\documentclass[doublecol]{epl2} 

\usepackage{amsmath}
\usepackage{amssymb}
\usepackage{graphicx}
\usepackage{dcolumn}
\usepackage{bm}
\usepackage{setspace}
\usepackage{amsfonts}
\usepackage{rotating}
\usepackage{makeidx}
\usepackage{amsthm}
\usepackage{color}

\newcommand{\openone}{\mathbb{I}}

\newcommand{\tr}{\mbox{Tr}}

\newcommand{\Diag}{\mbox{Diag}}

\newcommand{\sgn}{\mbox{sgn}}

\title{Local-hidden-state models  for T-states using finite  {shared randomness}
\footnote{EPL, 127 (2019) 20007}}
\shorttitle{Local-hidden-state models  for T-states using  {shared randomness} } 

\author{Yuan-Yuan Zhang \and Fu-Lin Zhang\footnote{Corresponding author: flzhang@tju.edu.cn} }
\shortauthor{Yuan-Yuan Zhang  and Fu-Lin Zhang}

\institute{
 Department of Physics, School of Science, Tianjin University - Tianjin 300072, China
}


\abstract{
The study of local models using finite shared randomness originates from the consideration about the cost of classically simulating entanglement in composite quantum systems.
We construct explicitly two families of local-hidden-state (LHS) models for T-states, by mapping the problem to the Werner state.
The continuous decreasing of shared randomness along with entanglement, as the anisotropy increases, can be observed in the one from the most economical model  for the Werner state.
The construction of the one for separable states shows that the separable boundary of  T-states can be generated from the one of the Werner state, and the cost is $2$ classical bits.
}

\begin{document}

\maketitle

\section{Introduction}

Nonclassical correlations in composite quantum systems and their hierarchy are fundamental issues in quantum information \cite{Book,RevModPhys.81.865,RMP2012Vedral,RMP2014bell,JPA2014LHV}.
Many concepts of these correlations can be traced back to the early days of quantum mechanics, and play key roles in several quantum information processes.
On the other hand, the tasks in quantum information also provide points of view to study the correlations.
An important example is the work of Wiseman  {\textit{et al.}} \cite{PRL2007Steering}, in which they define Bell nonlocality and Einstein-Podolsky-Rosen (EPR) steering according to two tasks, and prove that the former is a sufficient condition for the  {latter} and  entanglement is necessary for both of them.

In the tasks of Wiseman  {\textit{et al.}} \cite{PRL2007Steering}, two observers, Alice and Bob, share a bipartite entangled state.
Alice can affect the postmeasured states left to Bob by choosing different measurements on her half.
Such ability is  termed \textit{steering} by Schr\"{o}dinger \cite{steer1935}.
EPR steering from Alice to Bob exists when Alice can convince Bob that she has such ability, which is equivalent to the fact that unnormalized postmeasured states can not be described by a local-hidden-state (LHS) model.
Further, their state is Bell nonlocal, when the two observers can convince Charlie, a third person, that the state is entangled.
This is demonstrated by the inexistence of  local-hidden-variable (LHV) model explaining correlations of outcomes of their joint local measurement.
A LHS model is a particular case of a LHV model,  of which  the hidden variable is a single-particle state and one of the response functions is the probability of measurement on the state.

Construction of local models, especially the optimal ones, provides a division between the quantum and classical worlds, in the sense of whether the nonclassical correlations exist.
However,  it is an extremely difficult problem to explicitly derive optimal models.
Only a few models beyond Werner's results \cite{Werner1989} have been reported,  such as the ones in \cite{PRL2014oneway,PRL2005LHVBit,JOSAB2015Steering,arxiv2015UnSteer}, most of which are for states with high symmetries.
Our recent work \cite{zhang2017LHS} shows the possibility of generating local models for states with a lower symmetry, from the ones with a high symmetry. Namely, we obtain the optimal  {models for T-states (Bell diagonal states)}, given by Jevtic  {\textit{et al.}} \cite{JOSAB2015Steering} based on the steering ellipsoid \cite{PRL2014Ellips}, by mapping the problem to the one of the Werner state.

On the other hand, Bowles  {\textit{et al.}} \cite{PRL2005LHVBit} raise the issue of constructing local models using finite shared randomness.
This comes from their consideration about the cost of classically, measured by classical bits encoding the local variable, simulating the correlations in an entangled state.
They give a series of  LHV models for Werner states using finite shared randomness, and prove the existence of the ones for entangled states admitting a LHV model.
 {
These results inspire a method for constructing LHV models for entangled states, in which the problem of finding a local model for an infinite set of measurements is mapped to the one of a finite set of measurements \cite{Arxiv2015LHV,PRL2016Algorithmic,RPP2017RevLHS}.
In addition,  the concepts  of  superlocality  \cite{PRA2015SupLocal,PRA2017SupLocal} and superunsteerability \cite{PRA2018SupLHS} stemm from  the study of  shared classical randomness required to simulate local correlations.
}



In the present work, we
  {
  study local models for T-states by extending our strategy in \cite{zhang2017LHS}  to
 }
 the case with finite shared randomness.
 {
They are LHS models, as the shared local variables are sets of  discrete states on the Bloch sphere and Bob's response function is his measurement probability on these states.
Expressing the discrete distributions for Werner states in \emph{Protocols 1} and \emph{2} of \cite{PRL2005LHVBit}  in terms of Dirac delta functions, we derive a family of LHS models for  T-states by using the  mapping in \cite{zhang2017LHS}.
The one generated from the most economical LHV model  for Werner state is discussed in detail, which provides an example to observe the continuously changing  shared randomness with entanglement.
 Besides, we construct a LHS model,  not belonging to the two protocols in \cite{PRL2005LHVBit},  for the critical separable Werner state, by decomposing it into product states.
 It can be transformed into the LHS models for the critical separable T-states by a generalization of  the original mapping in \cite{zhang2017LHS}.
 This
 }
  shows the possibility of generating the separable boundary for a class of states with a low symmetry, and decomposing them into product states, from a higher symmetric case.



%

\section{Preliminaries}\label{LHSReview}

\subsection{LHS model}

We first give a brief review of the concepts of EPR steering and LHS model, under the context of two-qubit system and projective measurements.
An arbitrary two-qubit  state shared by Alice and Bob can be written as
\begin{equation}\label{rhoab}
{\rho}_{AB}=\frac{1}{4} (\openone \otimes \openone +\vec{a} \cdot \vec{\sigma} \otimes \openone +\openone \otimes \vec{b} \cdot \vec{\sigma} + \sum_{ij} T_{ij} \vec{\sigma}_{i} \otimes \vec{\sigma}_{j} ),
\end{equation}
where $ \openone$ is the unit operator, $ \vec{a} $ and $ \vec{b} $ are the Bloch vectors for Alice and Bob's qubit, $ \vec{\sigma} = ( \sigma_x,\sigma_y,\sigma_z ) $ is the vector of the Pauli operators, and $ T_{ij} $ is correlation matrix.
We focus on the case in which Alice makes a projection measurement on her part.
The measurement operator of Alice uniquely corresponds to a unit vector $\vec{x}$ and a outcome $a=\pm1$ as
\begin{equation}
 \Pi _a^{\vec{x}} = \frac{1}{2} ({\openone+ a\vec{x}\cdot \vec{\sigma} } ) .
 \end{equation}
After the measurement, Bob's state becomes
	\begin{equation} \label{ assem}
		\begin{split}
		\rho _a^{\vec{x}} &= \tr ( {\Pi _a^{\vec{x}} \otimes \openone{\rho _{AB}}} ) \\
		&= \frac{1}{4} [ { ( {1 + a\vec{a} \cdot \vec{x}} ) \openone + ( {\vec{b} + a{T^{\rm T}}\vec{x} } ) \cdot \vec{\sigma} }  ],
		\end{split}
\end{equation}
where $T^{\rm T}$ is transposed $T$.
The set of $\rho _a^{\vec{x}} $ is called an assemblage.

A LHS model is defined as
\begin{equation}\label{LHS}
 \rho^{\rm{LHS}} = \int \omega ( \vec{\lambda} )  p ( a| \vec{x},\vec{\lambda} ) \rho_{\vec{\lambda}} d \vec{\lambda} .
\end{equation}
Here, $ \rho_{\vec{\lambda}} $ is a hidden state depending on the hidden variable $ \vec{\lambda} $ with the distribution function $ \omega ( \vec{\lambda} ) $.
And, $ p ( a| \vec{x},\vec{\lambda} ) $ is a response function simulating the probability of Alice's outcome,  with $ p ( a| \vec{x},\vec{\lambda} ) \geqslant 0$ and $p ( 1| \vec{x},\vec{\lambda} ) + p ( -1| \vec{x},\vec{\lambda} ) = 1 $.
If there exists a LHS model satisfying
\begin{eqnarray}\label{AssEqLHS}
\rho _a^{\vec{x}} = \rho^{\rm{LHS}}
\end{eqnarray}
for all the measurements, the outcomes of Alice's measurements and Bob's collapsed state  can be simulated by a LHS strategy without any entangled state \cite{PRL2007Steering}.
On the contrary, if a LHS model satisfying (\ref{AssEqLHS}) does not exist, $\rho_{AB}$ is termed steerable from Alice to Bob.

 {Without loss of generality, we may} take a hidden variable to the unit Bloch vectors and the local hidden states to be corresponding pure qubit states {\cite{arxiv2015UnSteer}} as
\begin{equation}
\rho_{\vec{\lambda}} = |\vec{\lambda}\rangle \langle \vec{\lambda} |=\dfrac{1}{2} ( \openone + \vec{\lambda} \cdot \vec{\sigma} ) .
\end{equation}
Then, $d \vec{\lambda}$  is the surface element on  the Bloch sphere.
We can take
 \begin{equation}\label{PA}
  p ( a| \vec{x},\vec{\lambda} )  = \frac{1}{2}\bigr[ 1 + af(\vec{x},\vec{\lambda})\bigr] ,
  \end{equation}
  with $ f(\vec{x},\vec{\lambda}) \in [-1,1] $.
The LHS model can be rewritten as
\begin{equation}
{\rho ^{\rm LHS}} \!\!=\!\! \int \!\! { \omega ( \vec{\lambda} ) \!  \frac{1}{4} \biggr[  {\openone \! +\! \vec{\lambda}    \! \cdot \!  \vec{\sigma} \!+\! af(\vec{x},\! \vec{\lambda}) \!+\! a f(\vec{x},\! \vec{\lambda}) \vec{\lambda}  \! \cdot \! \vec{\sigma} }  \biggr] }     d\vec{\lambda}.
\end{equation}
Consequently, the equation (\ref{AssEqLHS}) requires
\begin{subequations}\label{Reqs}
\begin{align}
&\int \omega(\vec{\lambda}) d \vec{\lambda} =1 , \label{omega1}\\
&\int \omega(\vec{\lambda}) f (\vec{x},\vec{\lambda}) d \vec{\lambda} = \vec{a}\cdot\vec{x} , \label{a0} \\
&\int \omega(\vec{\lambda}) \vec{\lambda} d \vec{\lambda} =\vec{b} , \label{b0} \\
&\int \omega(\vec{\lambda})  f (\vec{x},\vec{\lambda}) \vec{\lambda} d \vec{\lambda} = T^{\rm T} \vec{x}. \label{Tx}
\end{align}
\end{subequations}
The spin correlation matrix can always be diagonalized by local unitary operations, which preserve steerability or unsteerability.
Hence, we consider the diagonalized $T$, that $T=\Diag \{T_x,T_y,T_z\}$, and omit its superscript ${\rm T}$ in the following parts of this article.
 {
Constructing
}
a LHS model for a state $\rho_{AB}$ is equivalent to finding a pair of $ \omega ( \vec{\lambda} )$ and $ f ( \vec{x} ,\vec{\lambda} )$ fulfilling these requirements.

\subsection{T-states}

The state  (\ref{rhoab}) is called a T-state, when the Bloch vectors, $\vec{a}$ and $\vec{b}$, vanish.
In our recent work \cite{zhang2017LHS}, we present an approach to derive the optimal LHS model for T-states.
We first assume the correlation matrix on the EPR-steerable boundary being $T_0$ and $T= t T_0$  with $t\geq0$.
That is, the T-state with $t>1$ is EPR-steerable, and the one with $0\leq t\leq 1$ admits a LHS model.

The key step is multiplying both sides of Eq. (\ref{Tx}) by $ T^{-1}_0 $ and defining the unit vector $ \vec{\lambda '}  =  {{T_0^{ - 1}\vec{\lambda}  }}{ |  {T_0^{ - 1}\vec{\lambda}  } |^{-1} } $,
where $ |\cdot| $ is the Euclidean vector norm.
 Then the condition (\ref{Tx}) is rewritten as
\begin{equation}\label{Tx1}
		\int \omega' ( \vec{\lambda}' )  \frac{1}{ | T_0 \vec{\lambda}' | } f (\vec{x},\vec{\lambda}) \vec{\lambda}' d\vec{\lambda}' = t \vec{x},
\end{equation}
where $ \omega' ( \vec{\lambda}' ) $ is the distribution function of the new defined hidden variable $\vec{\lambda}' $,
 and $ d\vec{\lambda}' $ is a surface element on its unit sphere.
These variables are connected by a Jacobian determinant as
\begin{equation}
d \vec{\lambda} = | \det T_0 | |  {T_0^{ - 1}\vec{\lambda}  } |^3 d \vec{\lambda}' , \ \ \
\omega ( \vec{\lambda} ) d\vec{\lambda} =\omega' ( \vec{\lambda}' ) d\vec{\lambda}'.
\end{equation}

 In the optimal LHS model of the critical Werner state \cite{PRL2007Steering,Werner1989}, with $T_0=- \Diag [1/2,1/2,1/2]$, the functions in (\ref{Tx1}) satisfy
\begin{equation}
\omega' ( \vec{\lambda}' )    =\frac{1}{2 \pi} | T_0 \vec{\lambda}' |, \ \ \ f (\vec{x},\vec{\lambda}) =\sgn (\vec{x}\cdot\vec{\lambda}').
\end{equation}
 We find that, these relations give exactly the optimal LHS model for an arbitrary T-state and leads to the critical condition \cite{zhang2017LHS,JOSAB2015Steering,EPL2016Tstate}
 \begin{equation}
		\int \frac{1}{2 \pi} | T_0 \vec{\lambda}' | d\vec{\lambda}' = 1.
\end{equation}
An explicit expression for this integral can be found in  the work of Jevtic \textit{et al.} \cite{JOSAB2015Steering}.

\section{ LHS models for T-States with finite  {shared randomness}}

We now generate the LHS models for  {T-states} with finite  {shared randomness}, using our approach reviewed above.
The formulas in the above section are derived based on continuous local variables.
To utilize these results, we represent the distribution of the finite hidden variables as delta functions.
We mainly concentrate on the details of the two models corresponding to the most economical one of the Werner state and the one for the separable Werner state.

\subsection{LHS model on the icosahedron}

In the most economical model simulating an entangled Werner state, Alice and Bob share  $i =\{1,...,12\}$ uniformly distributed,
corresponding to  $12$ vertices of the icosahedron represented by the normalized vectors $\vec{v}_i$.
That is, the distribution is given by
 \begin{equation}
 \omega ( \vec{\lambda} )  =   \sum_{i} \frac{1}{12} \delta ( \vec{\lambda} -\vec{v}_i ) .
  \end{equation}
The radius of a sphere inscribed inside the icosahedron is $l=\sqrt{(5+2\sqrt{5})/15}$.
The icosahedron is symmetric under $\vec{v}_i \rightarrow -\vec{v}_i $, and its vertices satisfy the  properties $\sum_{j} \sgn(\vec{v}_j \cdot \vec{v}_i) \vec{v}_j= 2 \gamma \vec{v}_i$ with $\gamma=1+\sqrt{5}$.
Then, the  vector $l\vec{x}$ can always be represented as a convex decomposition $l\vec{x}=\sum_i \omega_i \vec{v}_i $ with $\omega_i \geq0$ and $\sum_i \omega_i =1$.
Defining the function
 \begin{equation}
  f (\vec{x},\vec{\lambda}) = - \sum_i \omega_i \sgn ( \vec{v}_i \cdot \vec{\lambda} ) ,
 \end{equation}	
 one can obtain
  \begin{equation}
\int \! \biggr[\! \sum_{j} \frac{1}{12} \delta ( \vec{\lambda} -\vec{v}_j ) \! \biggr]\biggr [\! - \! \sum_i \omega_i \sgn ( \vec{v}_i \cdot \vec{\lambda} ) \! \biggr] \! \vec{\lambda}   d\vec{\lambda} = -\! \frac{\gamma l}{6} \vec{x},
\end{equation}
which is the relation (\ref{Tx}) for the Werner state.

 To fulfill the condition (\ref{Tx}), equivalently the equation (\ref{Tx1}),  for T-states, an intuitive construction is given by
\begin{eqnarray}
&& \omega' ( \vec{\lambda}' )   =  \sum_{i} \frac{S}{12} \delta ( \vec{\lambda}' - \vec{v}_{i} )  | T_0 \vec{\lambda}' |, \\
 &&  f (\vec{x},\vec{\lambda}) =  \sum_i \omega_i \sgn ( \vec{v}_{i} \cdot \vec{\lambda}' ), \label{fT}
\end{eqnarray}
where $S$ is a constant determined by the normalization condition (\ref{omega1}).
They  {lead} to the visibility parameter in (\ref{Tx1}) being
\begin{equation}\label{tLHS}
t = S \frac{\gamma l}{6} =\frac{2 \gamma l }{ \sum_i | T_0 \vec{ v}_{i} |  }.
\end{equation}
Straightforward calculation gives the distribution of $\vec{\lambda}$ as
\begin{equation}\label{omegaT}
	 	\omega (\vec{\lambda}) =  \sum_{i}\frac{S}{ 12  }   \delta ( \vec{\lambda}- { T_0 \vec{v}_{i} }{|T_0 \vec{v}_{i}|^{-1}} ) |T_0 \vec{v}_{i}|.
\end{equation}
Then, both the integrals in (\ref{a0}) and (\ref{b0}) can be easily checked to be zero, by using the symmetries $\omega (-\vec{\lambda}) =\omega (\vec{\lambda}) $ and $f (\vec{x},-\vec{\lambda}) =-f (\vec{x},\vec{\lambda}) $.
Therefore, the functions (\ref{fT}) and (\ref{omegaT}) represent a LHS model for the T-state with the visibility parameter in (\ref{tLHS}).
And the extension to smaller amounts of $t$ is straightforward.

Obviously, in our LHS models for T-states, the  hidden variable $i=\{1,...,12\}$  distributes nonuniformly, whose probability is proportional to $|T_0 \vec{v}_{i}|$.
The corresponding unit vectors $\vec{\lambda}'$ locate on $ \vec{v}_{i} $, and  the Bloch vector of hidden states $\vec{\lambda}$  on $ { T_0 \vec{v}_{i} }{|T_0 \vec{v}_{i}|^{-1}}$.
Both the distribution and visibility parameter, given in (\ref{tLHS}), covered by the model, depend on the orientation of the icosahedron.
A natural question is which orientation is optimal, in the sense of maximizing the parameter  $t$, or equivalently $S$.

\subsection{Optimal icosahedron}

Since it is a complex problem to perform general maximisations, we consider the special case with an axial symmetry that $|T_{0,x}|=|T_{0,y}|$ with the aid of numerical calculation.
Then, the relation between $|T_{0,x}|$ and $|T_{0,z}|$ can be written as a simple formula \cite{JOSAB2015Steering}.
And the orientation of the icosahedron can be represented by the intersection of Z-axis with the surface of the icosahedron.
There are three types of special points on the surface, which are vertices, midpoints of edges and centre of faces.
We suspect that the maximum of $S$ occurs at these special points.

\begin{figure}
 \includegraphics[width=6cm]{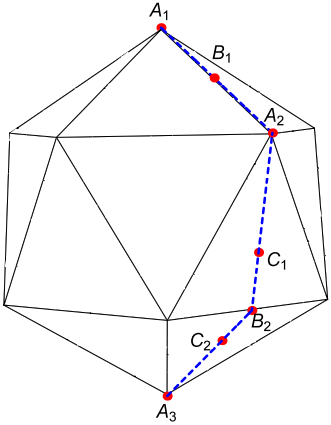} \centering
 \caption{ (Color online)
 The icosahedron in the construction of LHS models in \emph{Protocol 1}.
 Dashed blue lines show intersections of Z-axis with the surface of icosahedron during our rotation.
 Vertices, midpoints of edges and centre of faces on the dashed blue lines are marked as $A_i$, $B_j$ and $C_k$ respectively.
 }\label{icos}
\end{figure}

\begin{figure}
 \begin{flushright}
 \includegraphics[width=8.34cm]{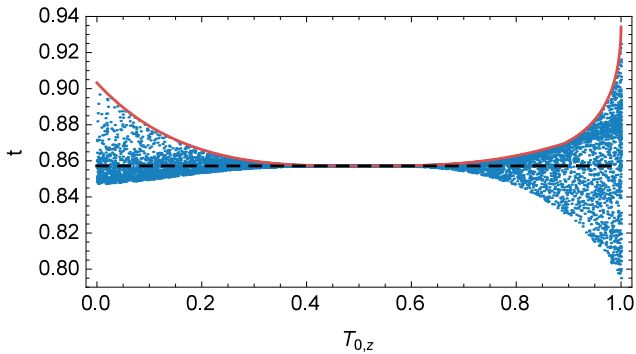}
 \end{flushright}
 \caption{ (Color online)
  The solid curve shows the maximum of $t$ in the LHS models based on the icosahedron, in company with the values for ten thousand random orientation, and the dashed line is for the value of the Werner state.
     }\label{figt}
\end{figure}

\begin{figure}
 \begin{flushright}
 \includegraphics[width=8cm]{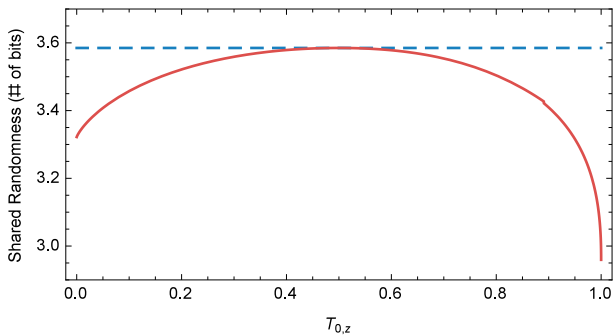}
 \end{flushright}
 \caption{ (Color online)
 The solid curve shows the shared randomness in the LHS model based on the icosahedron in optimal orientation, and the dashed line is for the value of the Werner state.
 }\label{figEntr}
\end{figure}

\begin{figure}[!h]
 \begin{flushright}
 \includegraphics[width=8.35cm]{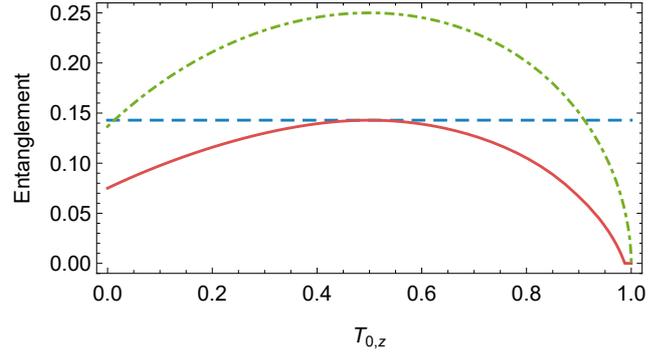}
 \end{flushright}
\caption{ (Color online)
The solid curve shows entanglement of T-states admitting  the optimal LHS model based on the icosahedron,
the dashed line is for the value of the Werner state,
and the  dot-dashed curve is for T-states on the EPR-steerable boundary.
}\label{figEntangle}
\end{figure}

Choosing a trajectory of the intersection, as shown in Fig. \ref{icos},  consisting of an edge and two medians of faces, one can plot the values of $S$ versus the location of intersection (we omit the figures here).
These curves indicate that the maximum of $S$ on the trajectory occurs at vertices when $|T_{0,z}|\leq 1/2$ , at the centre of faces when $1/2<|T_{0,z}| \lesssim 0.89$,
or at midpoints of edges  {when} $ |T_{0,z}| \gtrsim 0.89$.
These maximums, in the same sequence, can be analytically expressed as
\begin{eqnarray}
&& S_A = \frac{6}{  \sqrt{Z} +    \sqrt{  20 X     + 5 Z } }, \\
&& S_C = \frac{ \sqrt{30}} {  \sqrt{X  {\alpha_+}   + Z   {\beta_-}   }  +  \sqrt{X   {\alpha_- }   + Z  {\beta_+}   }},\\
&& S_B = \frac{3\sqrt{10}}{\sqrt{ 10X}  + \sqrt{ X {\alpha_+}+ Z {\alpha_-} } + \sqrt{ X   {\alpha_-} +Z  {\alpha_+} } }, \ \ \ \ \ \
\end{eqnarray}
 where $X=T_{0,x}^2$, $Z=T_{0,z}^2$, $\alpha_{\pm}=5 \pm \sqrt{5}$ and $\beta_{\pm}=5/2 \pm \sqrt{5} $.
We find that they are optimal among arbitrary orientations of the icosahedron, by comparing them with one hundred thousand randomly generated intersections.
One-tenth of the random points are shown in Fig. \ref{figt}, in company with the corresponding maximums of $t$.

Our construction provides a family of LHS models with a fixed dimensionality of the local variable.
It is interesting to observe the continuously changing shared randomness, and its relation with the region of T-states admitting our models.
We plot the maximums of $t$ in Fig. \ref{figt}, which measure how close our models get to the EPR-steerable boundary.
The corresponding shared randomness, measured by the entropy of the distribution (\ref{omegaT}) \cite{PRL2005LHVBit},
 {
$H=-\sum_i q_i \log_2 q_i$ with $q_i= |T_0\vec{v}_i|S/12$,
}
 is shown in   Fig. \ref{figEntr}.
Obviously, the visibility parameter and entropy  show two opposite trends.
The anisotropy of the correlation matrix enhances the maximums of $t$, while it decreases the entropy.
Among the family of  T-states, the LHS model for the Werner state, with the maximum distance to the EPR-steerable boundary,  requires the most shared randomness.

This anomalous phenomenon prompts us to go back to the original point: the cost of classically simulating the correlations of entangled states \cite{PRL2005LHVBit}.
It is direct to derive the entanglement,  {measured} by concurrence \cite{Wootters98}, for axial symmetric T-states as $\max \{0, (2t|T_{0,x}|+t|T_{0,z}|-1)/2\}$.
As shown by the solid line in Fig. \ref{figEntangle},  the entanglement reveals a similar tend as the number of classical bits to  {simulate} it.
The degree of entanglement reaches its maximum at the point of Werner states, $T_{0,z}=1/2$, and decreases with the anisotropy.
Comparing with the maximums of $t$, one can find that the points on the EPR-steerable boundary with small entanglement are easy to approach, in the sense of the cost of classically simulating the correlations of entangled states.

In Fig. \ref{figEntangle}, a noteworthy point  is the small interval with zero entanglement.
This indicates that our LHS models are not the most economical ones, at least for the separable state in the small interval.
This is because a $4$-dimensional local variable is sufficient to simulate a separable two-qubit state, while the least shared bits in our construction for T-states is $2.96$.
We shall present more discussion about the LHS models for separable T-states below.

\subsection{Separable boundary}

The above results can be straightforwardly extended to any LHS model for the Werner state in \emph{Protocols 1 and 2} of \cite{PRL2005LHVBit} using  a $3$-dimensional polyhedron  with $D$ vertices.
We omit these formulas for brevity.

In this part, we focus on the case with a shared variable of  dimension $D=4$.
In the results for Werner state \cite{PRL2005LHVBit}, the tetrahedron, with $4$ vertices,  is without inversion symmetry and hence be excluded from \emph{Protocols 1 and 2}.
On the other hand, the maximum visibility parameter one can simulate with  $D=4$ is the boundary of the separable Werner state \cite{PRL2005LHVBit}.
Here our question is \emph{whether one can generate the boundary of the separable T-states from the one of the Werner state, as we do in the study of EPR-steering \cite{zhang2017LHS}}.

To answer the above question, we restrict the response function to the form
\begin{equation}
 f(\vec{x},\vec{\lambda})=\vec{x}\cdot \vec{\eta} ,
\end{equation}
where $\vec{\eta} $ is Alice's Bloch vector depending on $\vec{\lambda}$.
We term the corresponding LHS model as a \emph{LHS model for separable state}.
The entanglement of  the two-qubit state $\rho_{AB}$ is demonstrated by the inexistence of a LHS model with   $ f(\vec{x},\vec{\lambda})$ in the above form.

On can derive the solution for Werner states to the conditions (\ref{Reqs}), by decomposing the critical separable Werner state into four product  states.
Let the Bell states $|\Psi_{\pm}\rangle=(|00\rangle\pm|11\rangle)/\sqrt{2}$ and $|\Phi_{\pm}\rangle=(|01\rangle\pm|10\rangle)/\sqrt{2}$.
The critical separable Werner state is $\rho_{AB}^w=(3 |\Phi_-\rangle \langle\Phi_-|+|\Phi_+\rangle \langle\Phi_+|+|\Psi_+\rangle \langle\Psi_+|+|\Psi_-\rangle \langle\Psi_-| )/6$.
We assume the normalized state
\begin{equation}
|\phi_i\rangle \propto \sqrt{3}|\Phi_-\rangle + e^{i \theta_{i1}} |\Phi_+\rangle + e^{i \theta_{i2} }|\Psi_+\rangle + e^{i \theta_{i3}} |\Psi_-\rangle,
 \end{equation}
to be separable, and to satisfy $\rho_{AB}^w=\sum^4_i |\phi_i\rangle\langle \phi_i|/4$.
There exist two solutions to these conditions, one of which is given by
\begin{eqnarray}
| \phi_1 \rangle \! =\! (\sin\! \frac{\alpha}{2}\! | 0  \rangle \!-\! \cos\! \frac{\alpha}{2}\!e^{i \beta}\!| 1   \rangle  )\!  \otimes \! (\cos\! \frac{\alpha}{2}\!| 0 \rangle \!+\! \sin\!\frac{\alpha}{2}\! e^{i\beta}\!| 1 \rangle),
\end{eqnarray}
  $| \phi_2 \rangle=\sigma_x\otimes\sigma_x | \phi_1 \rangle$,  $| \phi_3 \rangle=\sigma_y\otimes\sigma_y | \phi_1 \rangle$, and $| \phi_4 \rangle=\sigma_z\otimes\sigma_z | \phi_1 \rangle$ , where $\alpha=\arccos (1/\sqrt{3})$ and  $\beta=-{\pi}/{4}$.
Alice's measurements on the decomposition $\rho_{AB}^w=\sum^4_i |\phi_i\rangle\langle \phi_i|/4$ lead to a LHS model using the tetrahedron.
Namely, it is defined by
\begin{eqnarray}
\omega(\vec{\lambda})=\sum_{i} \frac{1}{4} \delta(\vec{\lambda}-\vec{v}_i) ,\ \ \  \vec{\eta} =-\vec{\lambda},
\end{eqnarray}
with the $4$ vertices of the tetrahedron $\vec{v}_1=(1,-1,1)/\sqrt{3}$, $\vec{v}_2=(1,1,-1)/\sqrt{3}$ ,  $\vec{v}_3=(-1,1,1)/\sqrt{3}$ and $\vec{v}_4=(-1,-1,-1)/\sqrt{3}$.
They satisfy
\begin{eqnarray}\label{SepW}
 \int \biggr[\sum_{i} \frac{1}{4} \delta(\vec{\lambda}-\vec{v}_i)\biggr] \biggr[  \vec{x}\cdot(-\vec{\lambda})\biggr] \vec{\lambda} d\vec{\lambda}=-\frac{1}{3}\vec{x}.
\end{eqnarray}
The other solution leads to a model on the mirror image of the tetrahedron.

 We now turn to the T-states on the separable boundary.
 It is  universal to consider a positive definite $T$, as any minus sign can be merged into $\vec{\eta}(\vec{\lambda})$.
 Here we perform $T^{-\frac{1}{2}}$ on the condition (\ref{Tx}), and  define the unit vector $ \vec{\lambda }''  =  {{T^{-\frac{1}{2}}\vec{\lambda}  }}{ | {T^{-\frac{1}{2}}\vec{\lambda}  } |^{-1} } $ and its distribution $\omega'' ( \vec{\lambda}'') $.
 Then the condition (\ref{Tx}) for a separable state is rewritten as
\begin{equation}\label{Tx3}
		\int \omega'' ( \vec{\lambda}'' )  \frac{1}{ | T^{\frac{1}{2}}\vec{\lambda}'' | }\bigr[(T^{-\frac{1}{2}} \vec{x})\cdot\vec{\eta} \bigr] \vec{\lambda}'' d\vec{\lambda}'' = \vec{x}.
\end{equation}
Defining the unit vector $ \vec{\eta }''  =  {{T^{-\frac{1}{2}}\vec{\eta}  }}{ | {T^{-\frac{1}{2}}\vec{\eta}  } |^{-1} } $,
 one can find that $ (T^{-\frac{1}{2}} \vec{x})\cdot\vec{\eta}    = \vec{x} \cdot  \vec{\eta }''  | T^{\frac{1}{2}}\vec{\eta}'' |^{-1}$.
From the integral (\ref{SepW}), it was easy to find a pair of $ \omega'' ( \vec{\lambda}'' )  $ and $\vec{\eta}''$ satisfying (\ref{Tx3}) as
 \begin{equation}
 \omega'' ( \vec{\lambda}'' ) =\sum_{i} \frac{3}{4} \delta(\vec{\lambda}''-\vec{v}_i)| T^{\frac{1}{2}}\vec{v}_i |^{2},\ \ \ \vec{\eta}''=\vec{\lambda}''.
\end{equation}
 The normalization condition (\ref{omega1}) and the coordinates of  {$\vec{v}_i$} lead to
 \begin{equation}
 | T_x|+|T_y|+|T_z|=1,
 \end{equation}
 which is nothing but the separable boundary  of T-states \cite{horodecki1996information,dakic2010necessary}.
 Then, in the space of $\lambda$, the distribution and Bloch vector of Alice are
  \begin{equation}\label{omegaTS}
 \omega ( \vec{\lambda}) =\sum_{i} \frac{1}{4} \delta\bigr(\vec{\lambda}-\sqrt{3}T^{\frac{1}{2}} \vec{v}_i\bigr),\ \ \ \vec{\eta}=\vec{\lambda}.
\end{equation}
Substituting them into the equations (\ref{a0}) and (\ref{b0}), one can confirm both the integrals to be zero.

In the LHS models for separable T-states, defined by  (\ref{omegaTS}), the shared variables are encoded on $\sqrt{3}T^{\frac{1}{2}} \vec{v}_i$, and are uniformly distributed.
The amount of shared randomness is $2$ bits, which is not affected by the  anisotropy of the correlation matrix.
The models are optimal in the sense of  reaching the separable boundary.
However, the question as to whether they are the most economical remains open.

\section{Summary}

We study LHS models for T-states using finite shared randomness.
The models are generated from the ones for Werner states, two of which are mainly discussed.
The first is derived by using our recent approach \cite{zhang2017LHS} on the most economical model  for the Werner state.
It provides an example to observe the continuously changing  shared randomness with an entangled state.
With the increase of anisotropy,  the amount of shared classical bits drops along with entanglement, although the model  gets closer to the EPR-steerable boundary.
The second one is restricted to simulate a separable state by a condition on Alice's response function.
 It is derived from the one for the Werner state by a generalized generating approach, and reaches exactly  the separable boundary of  T-states.
 The cost of classical randomness in this model is $2$ bits, which is not affected by the anisotropy of the correlation matrix.

It would be interesting to consider the open questions or extensions below.
First, our approach to derive the LHS models for T-states on the separable boundary is actually to decompose them into product states.
Geneneralizing this method may be a starting point to define T-states in higher-dimensional systems,  which has been raised in our recent work \cite{zhang2017LHS}.
Second, in what region our model using the icosahedron is the most economical one?
Third,  what is the minimal cost to classically simulate a separable state?
This is a nontrivial question, as in LHS models on the separable boundary, the amount of bits is different from the entropy of  states.
This difference originates from the superposition of states in composite quantum systems, and may be interpreted as a kind of quantum correlation.

\acknowledgments
This work is supported by the NSF of China (Grant No. 11675119, No. 11575125 and No. 11105097).

\bibliography{LHSTStateBits}

\end{document}